\documentclass[epj]{svjour}
\usepackage{graphicx}
\begin{document}
\title{
Electric dipole response of $^{208}$Pb from proton inelastic scattering:
constraints on neutron skin thickness and symmetry energy
}
\author{
A.~Tamii\inst{1}
\and
P.~von~Neumann-Cosel\inst{2}
\and
I.~Poltoratska\inst{2}
}                     
%
%
\institute{
Research Center for Nuclear Physics, 10-1 Mihogaoka, Ibaraki 567-0047, Japan
\and
Institut f\"{u}r Kernphysik, Technische Universit\"{a}t Darmstadt, D-64289 Darmstadt, Germany
}
\date{Received: date / Revised version: date}
%
\abstract{
The electric dipole ($E1$) response of $^{208}{\rm Pb}$ has been precisely determined by
measuring Coulomb excitation induced by proton scattering at very forward angles.
The electric dipole polarizability, defined as inverse energy-weighted sum rule of the $E1$ strength, 
has been extracted as $\alpha_D=20.1\pm0.6$ fm$^3$.
The data can be used to constrain the neutron skin thickness of $^{208}{\rm Pb}$ to 
$\Delta r_{np}=0.165\pm(0.009)_{\rm expt}\pm(0.013)_{\rm theor}\pm(0.021)_{\rm est}$ fm,
where the subscript ``expt'' refers to the experimental uncertainty, ``theor'' to the theoretical confidence
band and ``est'' to the uncertainty associated with the estimation of the symmetry energy
at the saturation density.
In addition, a constraint band has been extracted in the plane of the symmetry energy ($J$) 
and its slope parameter ($L$) at the saturation density.
} 
\maketitle

\section{Introduction}
\label{sec:intro}

The nuclear equation of state (EOS) defines the bulk properties of nuclear matter from atomic nuclei to neutron stars.
Determination of the nuclear EOS is one of the fundamental goals of nuclear physics.
Since nuclear matter is composed of two kinds of particles, {\it i.e.} neutrons and protons, the EOS contains a term -- the symmetry energy -- which depends on the density asymmetry between neutrons and protons.
Determination of the symmetry energy term currently draws much attention both theoretically and experimentally, as illustrated by this special issue.
An accurate determination of the symmetry energy allows precise predictions of properties of exotic nuclei with large differences between proton and neutron numbers.
The symmetry energy is furthermore a basic input for calculations of heavy-ion collision processes, where isospin-asymmetric matter is produced and leads to density-dependent reactions.
In astrophysics, the symmetry energy is relevant to the properties of neutron stars, such as mass, radius and internal structure, the supernova-explosion process, neutron star cooling, and other dynamical processes related to neutron rich matter.

The nuclear EOS can be studied in laboratory by measuring data on nuclear ground and excited states and constructing theoretical models which attempt to describe them.
In this article, we report on a precise determination of the electric dipole response of $^{208}{\rm Pb}$ by measuring relativistic Coulomb excitation induced by proton inelastic scattering at very forward angles.
The dipole polarizability, defined as the inverse energy-weighted sum rule of the electric dipole response, is closely related to the neutron skin thickness and to the density dependence of the symmetry energy.

\section{Relation between symmetry energy, neutron skin thickness and dipole polarizability}
\label{sec:rel}

The EOS of cold nuclear matter can be approximately written as a sum of the energy per nucleon of symmetric matter and an asymmetry term~\cite{Tsang2012}
\begin{equation}
\label{eq:eos}
E(\rho,\delta) = E(\rho,\delta=0)+S(\rho)\delta^2+O(\delta^4)\,,
\end{equation}
where the nucleon density ($\rho$) and the asymmetry parameter ($\delta$) are defined by the neutron ($\rho_n$) and proton ($\rho_p$) density as
\begin{eqnarray}
\rho & \equiv & \rho_n+\rho_p \,,\\
\delta & \equiv & \frac{\rho_n-\rho_p}{\rho_n+\rho_p}\,.
\end{eqnarray}
The symmetry energy factor $S(\rho)$ in Eq.~(\ref{eq:eos}) can be expanded around the saturation density $\rho_0\sim0.16$~fm$^{-3}$ as
\begin{equation}
\label{exp-sd}
S(\rho) = J+\frac{L}{3\rho_0}(\rho-\rho_0) 
+\frac{K_{\rm sym}}{18\rho_0^2}(\rho-\rho_0)^2
+\cdots\,.
\end{equation}
Here, $L$ is the slope parameter at density $\rho_0$.
It governs the pressure from the symmetry energy in pure neutron matter and the baryonic pressure in neutron stars~\cite{Horowitz2001}, and its value is approximately proportional to the fourth power of the neutron star radius~\cite{Lattimer2007}.

\begin{figure*}[t]
\begin{center}
\includegraphics[width=30pc]{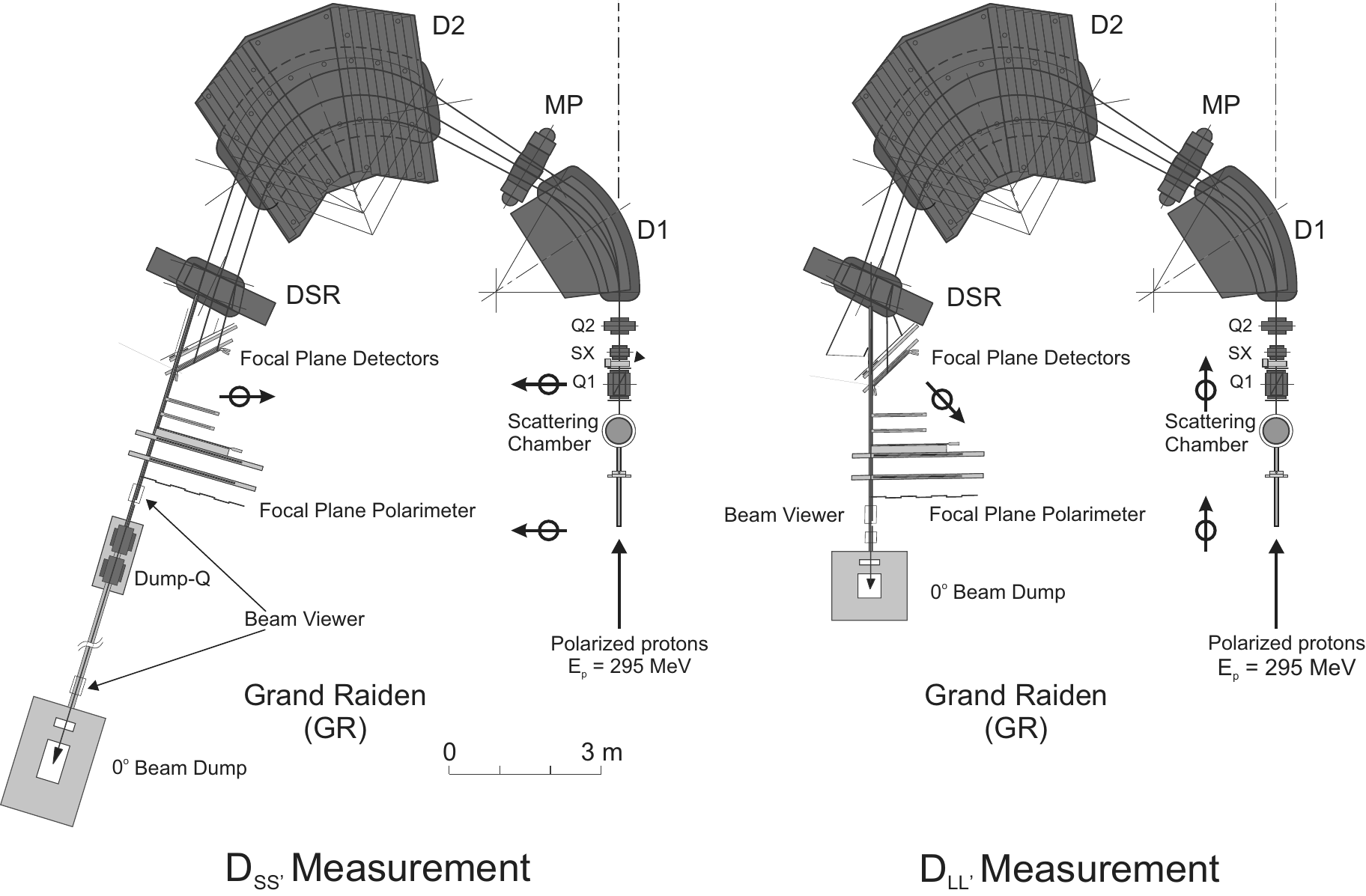}
\caption{\label{fig:setup}
Experimental setup of the Grand Raiden spectrometer at zero degree
for the measurement of the  $D_{SS'}$ (left) and $D_{LL'}$ (right) polarization transfer coefficients in inelastic proton scattering off $^{208}$Pb at $E_0 = 295$ MeV.
See text for details.
}
\end{center}
\end{figure*}

Linear correlations between the slope parameter and the neutron skin thickness of $^{208}{\rm Pb}$ are predicted by self-consistent mean field model calculations with various sets of interaction parameters~\cite{Roca-Maza2011,Piekarewicz2012}.
Here, the neutron skin thickness $\Delta r_{np}$ is defined as the difference of the root-mean-square radii of neutrons and protons.
Thus, experimental data on the neutron skin thickness may provide constraints on the slope parameter.

Parity-violating asymmetry measurements of electron elastic scattering (PREX experiment) at Jefferson Laboratory~\cite{PREX} are the most model-independent way to determine the neutron skin thickness of $^{208}{\rm Pb}$.
The electro-weak interaction is used to probe the form factor of the neutron density distribution.
However the uncertainty of the latest result, $\Delta r_{np}$=0.302 $\pm$ 0.175 $\pm$ 0.026 $\pm$ 0.005 fm~\cite{PREXResult}, is too large to impact on the allowed range of the slope parameter.
An improved measurement with better statistics is highly desired.

Reinhard and Nazarewicz reported a strong correlation between the dipole polarizability of $^{208}{\rm Pb}$ and its neutron skin thickness in the framework of a self-consistent mean field calculation based on the energy-density functional method with the SV$_{\rm min}$ Skyrme interaction \cite{Reinhard2010}.
The dipole polarizability $\alpha_D$ can be expressed as the inverse energy-weighted sum-rule of the electric dipole ($E1$) reduced transition probability $B(E1)$
\begin{equation}
\label{eq:dp}
\alpha_D = \frac{\hbar c}{2\pi^2}\int\frac{\sigma_{abs}}{\omega^2}d\omega
 = \frac{8\pi}{9}\int\frac{S_{E1}(\omega)}{\omega}d\omega\,,
\end{equation}
where $\omega$ stands for the excitation energy, $\sigma_{\rm abs}$ for the photo-absorption cross section, and $S_{E1}(\omega)=dB(E1)/d\omega$ for the $B(E1)$ strength per unit excitation energy.
The dipole polarizability can be experimentally determined by measuring the $B(E1)$ distribution as a function of the excitation energy.
We have used proton inelastic scattering at very forward angles to study the $B(E1)$ distribution of $^{208}{\rm Pb}$ by relativistic Coulomb excitation.

\section{Experimental Method}
\label{sec:exp}

The experiment has been performed at the Research Center for Nuclear Physics (RCNP), Osaka University.
Details of the experimental experimental method can be found in Ref.~\cite{Tamii2009} and details for the $^{208}$Pb experiment in Refs.~\cite{Tamii2011,Poltoratska2012}.
A polarized proton beam has been accelerated to 295 MeV with a beam intensity of 2-10~nA 
and a polarization degree of about 0.7.
An isotopically enriched ${}^{208}{\rm Pb}$ foil with a thickness of 5.2 mg/cm$^2$ has been used as a target. 
An energy resolution of 30~keV (FWHM) was achieved by dispersion matching techniques.
The experimental setup is shown in Fig.~\ref{fig:setup} for two different  polarization transfer measurements.
The primary beam passing through the target was transported into the {\it Grand Raiden} spectrometer~\cite{GR}, extracted at the focal plane, and  stopped in the beam dump.
The sideway-to-sideway polarization transfer $D_{SS'}$ (l.h.s .\ of Fig.~\ref{fig:setup}) was measured with a sideway polarized beam and the standard focal plane without use of the dipole spin-rotation (DSR) magnet.
The longitudinal-to-longitudinal polarization transfer $D_{LL'}$ (r.h.s .\ of Fig.~\ref{fig:setup}) was measured with a longitudinally polarized beam and rotation of the scattered protons by $18^\circ$ utilizing the DSR magnet.
Polarization transfer coefficients were measured in an angular range $0^\circ - 2.5^\circ$.
Differential cross sections were measured for angles between 0 and 10 degrees.
The excitation energy range covered by the momentum acceptance of the spectrometer was about $5-25$ MeV.

\section{Experimental results}
\label{sec:exp-results}

\begin{figure}[tbh]
\begin{center}
\includegraphics[width=20pc]{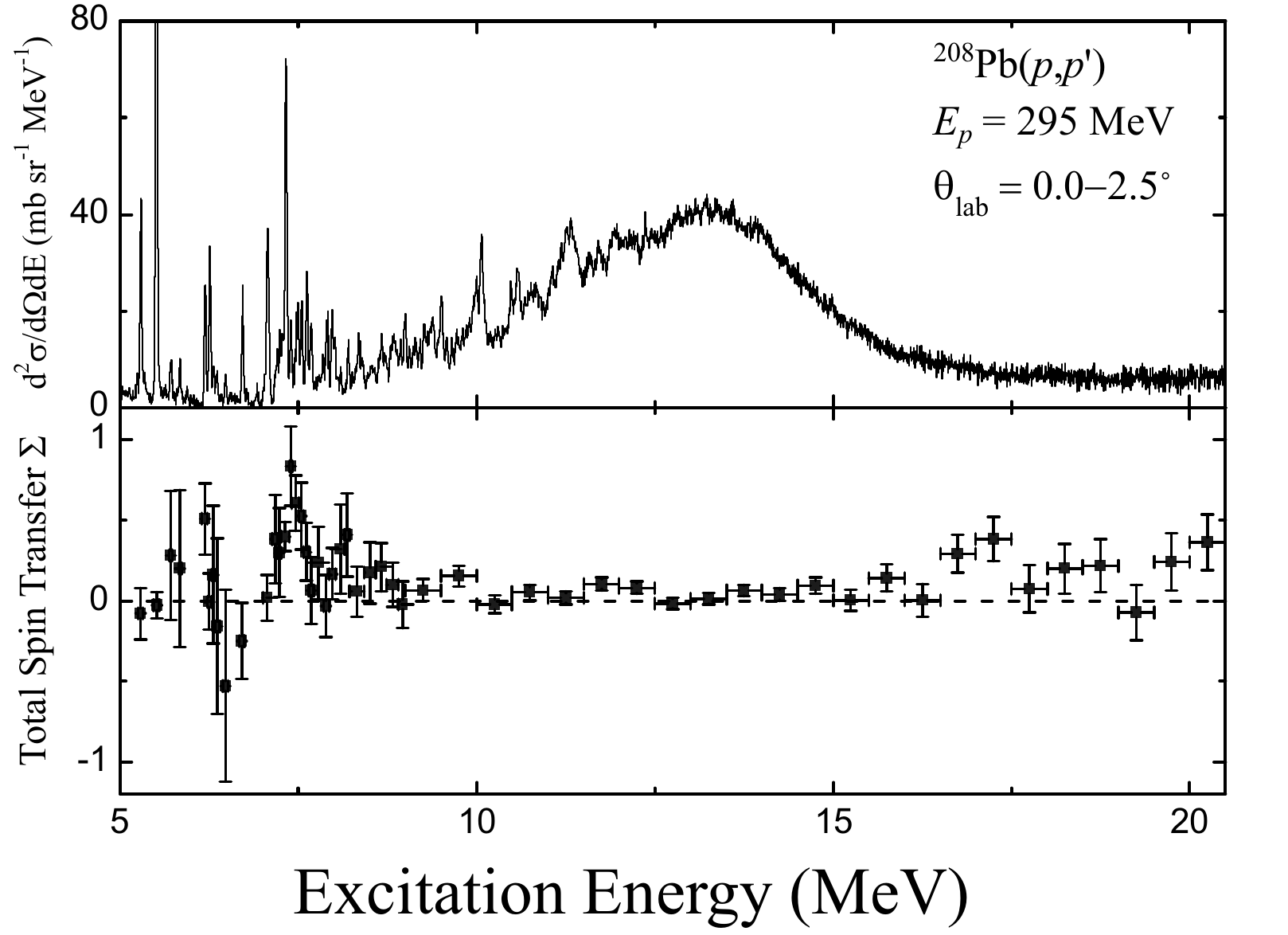}
\caption{\label{fig:DCSandTST}
(upper panel) Double differential cross sections and (lower panel) total spin transfer of the $^{208}{\rm Pb}$(p,p') reaction at $E_p$=295 MeV and at 0-2.5 degrees.
}
\end{center}
\end{figure}

The upper panel of Fig.~\ref{fig:DCSandTST} shows the double differential cross section of the $^{208}{\rm Pb}(p,p')$ reaction with the spectrometer set at $0^\circ$ and covering an angular range up to $2.5^\circ$.
The giant dipole resonance (GDR) is clearly visible as bump structure centered at about 13 MeV with fine structures on the lower-energy tail.
Discrete transitions were observed below the neutron separation energy ($S_n$=7.368 MeV).

\begin{figure}[tbh]
\begin{center}
\includegraphics[width=20pc]{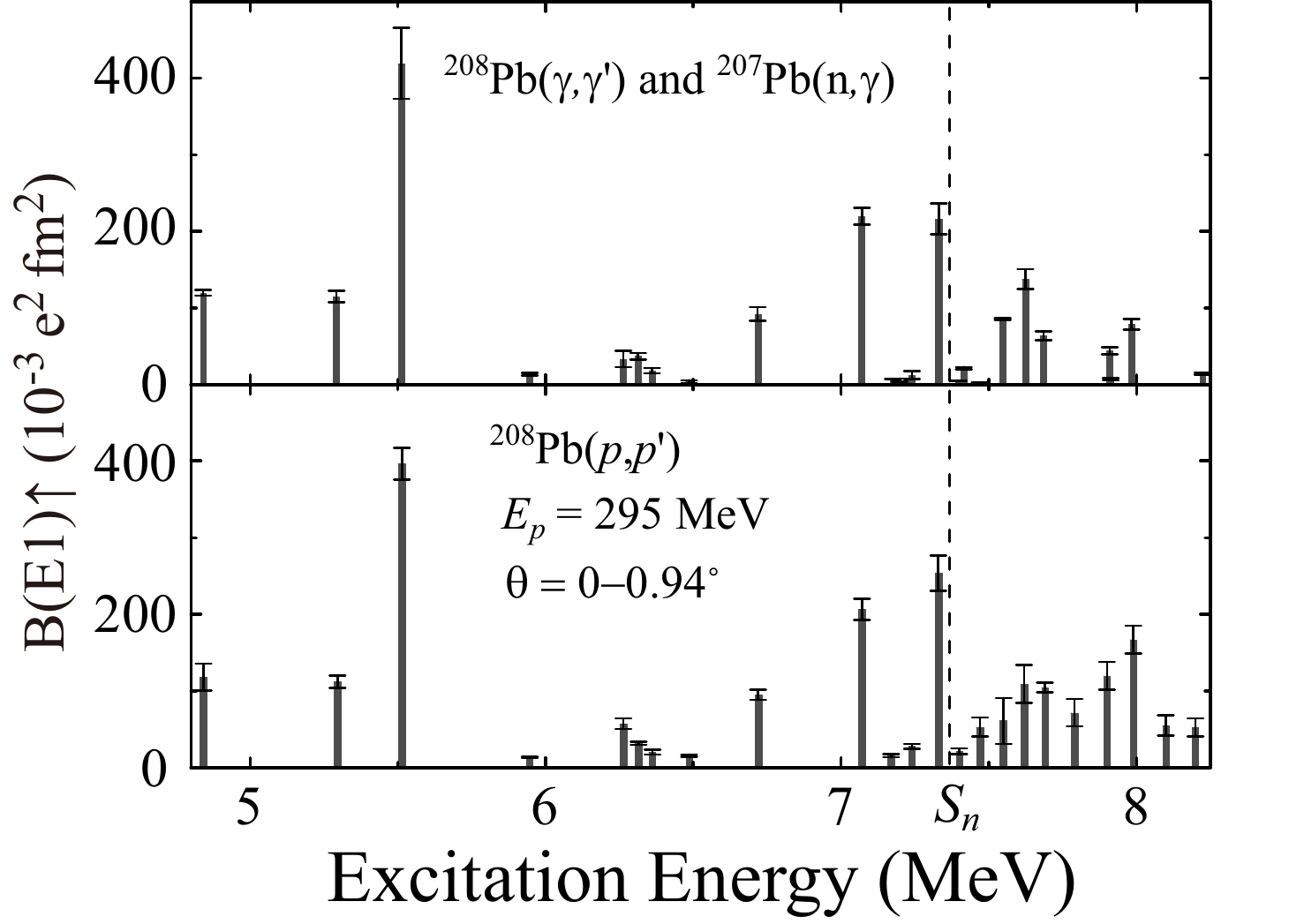}
\caption{\label{fig:DS}
Electric dipole reduced transition probability $B(E1)$ of low-lying discrete states 
determined by (upper panel) real-photon measurements~\cite{Ryezayeva2002,Enders2003,Shizuma2008,Schwengner2010}
 and (lower panel) the $(p,p')$ experiment.
}
\end{center}
\end{figure}

As demonstrated in Fig.~\ref{fig:DS}, there is a one-to-one correspondence with $E1$ transitions observed in nuclear resonance fluorescence (NRF) measurements \cite{Ryezayeva2002,Enders2003,Shizuma2008,Schwengner2010}.
For each low-lying discrete transition, the $B(E1)$ strength has been extracted from the $(p,p')$ cross sections in the angular range $0^\circ - 0.94^\circ$ assuming that it arises purely from Coulomb excitation.
The extracted $B(E1)$ values agree very well with the NRF data below $S_n$~\cite{Ryezayeva2002,Enders2003,Shizuma2008,Schwengner2010}.
Above $S_n$ more strength than previously known has been observed~\cite{Poltoratska2012}.

Since the experimental kinematics also favor excitation of the spin-$M1$ resonance, a decomposition of $E1$ and $M1$ cross sections is necessary for an extraction of the $E1$ strength in the continuum region.
This has been achieved with two independent methods.
The first method utilizes polarization transfer data.
One can define the total spin transfer $\Sigma$ 
\begin{equation}
  \Sigma \equiv \frac{3-(D_{SS'}+D_{NN'}+D_{LL'})}{4}\,,
\label{eq:sigma}
\end{equation}
where $D_{SS'}$, $D_{NN'}$ and $D_{LL'}$ are the sideway, normal and longitudinal spin transfer coefficients, respectively~\cite{Ohlsen1972}.
The total spin transfer $\Sigma$ becomes unity for spin-$M1$ and other spin-flip excitations by the nuclear interaction and zero for (Coulomb-excited) $E1$ and other non-spin-flip excitations.
Relation (\ref{eq:sigma}) can be derived at zero degrees from parity conservation~\cite{Suzuki2000} and thus holds model-independently.
Note that $D_{SS'} = D_{NN'}$ at zero degrees from rotational symmetry, and therefore only two polarization transfer coefficients need to be measured.
The resulting total spin transfer for $^{208}$Pb is plotted in the lower panel of Fig.~\ref{fig:DCSandTST}.
The GDR bump region is dominantly composed of $E1$ strength.
A concentration of spin-$M1$ strength is observed in the $7-8$ MeV region consistent with polarized real-photon measurements~\cite{Shizuma2008,Laszewski1998}.

\begin{figure}[tbh]
\begin{center}
\includegraphics[width=20pc]{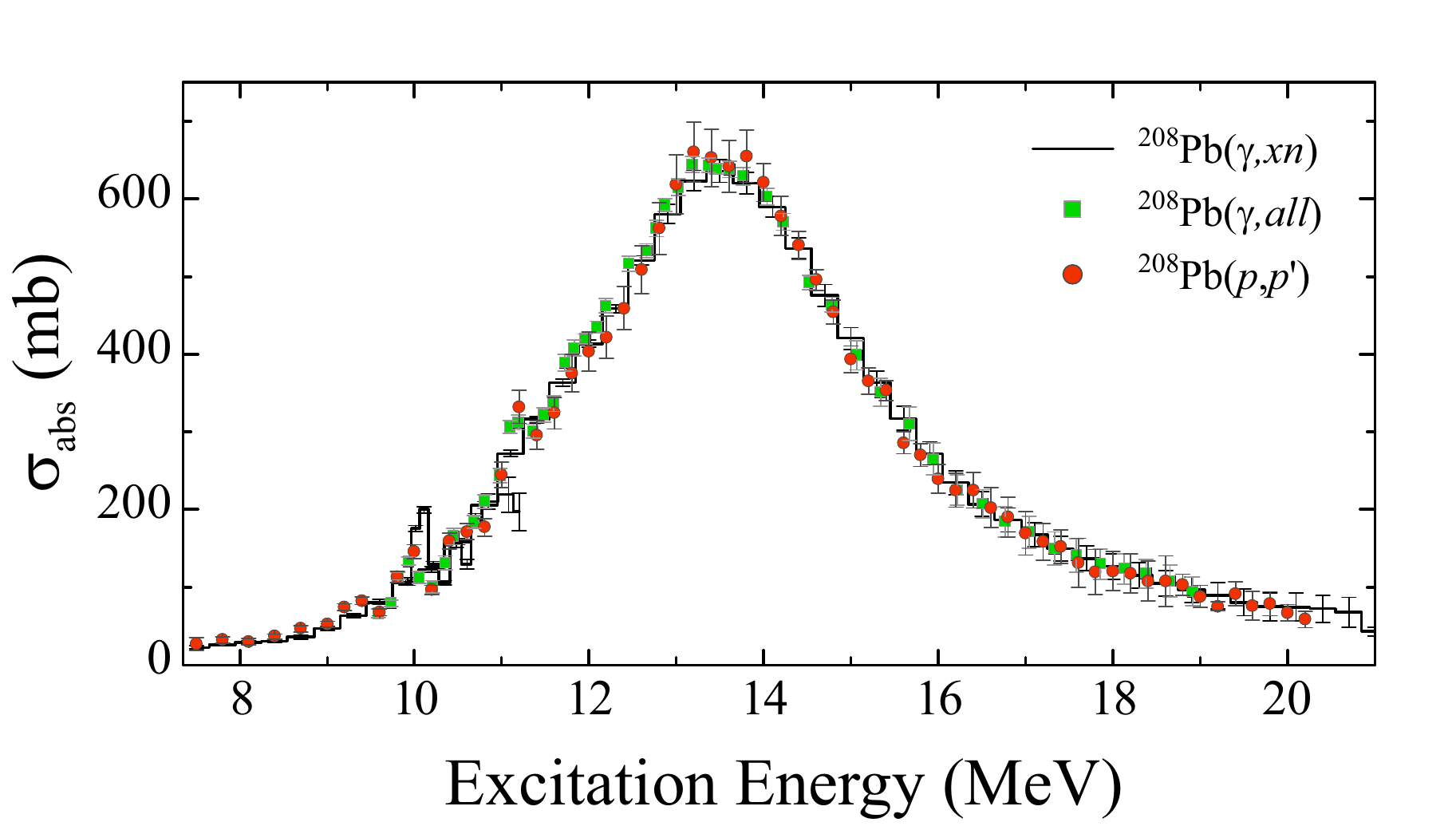}
\caption{\label{fig:GDR}
Comparison of photo-absorption cross sections determined by $(p,p')$ (red circles),  $(\gamma,xn)$~\cite{Veyssiere1970} (histogram) and tagged gamma-absorption~\cite{Schelhaas1988} (green squares) experiments in the GDR region of $^{208}{\rm Pb}$.
}
\end{center}
\end{figure}

The second method is a multipole-decomposition analysis (MDA) utilizing the angular distributions of the cross sections in each excitation energy bin to decompose contributions from different multipolarities.
Angular distribution shapes for each multipolarity were calculated in the distorted wave impulse approximation (DWIA) with the code DWBA07~\cite{DWBA07} and the effective interaction of Franey and Love ~\cite{Franey1985}.
RPA amplitudes and single-particle wave functions were taken from quasi-particle phonon model (QPM) calculations~\cite{Ryezayeva2002}.
For each energy bin, the contributions of different multipolarities were determined by a least-square fit reproducing the experimental angular distribution. 
The results of the two methods agree to each other within error bars~\cite{Tamii2011}.
The $E1$ photo-absorption cross section in the GDR region extracted by the MDA is shown in Fig.~\ref{fig:GDR} as red circles.
The result is consistent with $(\gamma,xn)$~\cite{Veyssiere1970} (black histogram) and tagged-photon~\cite{Schelhaas1988} (green squares) measurements.

\begin{figure}[tbh]
\begin{center}
\includegraphics[width=20pc]{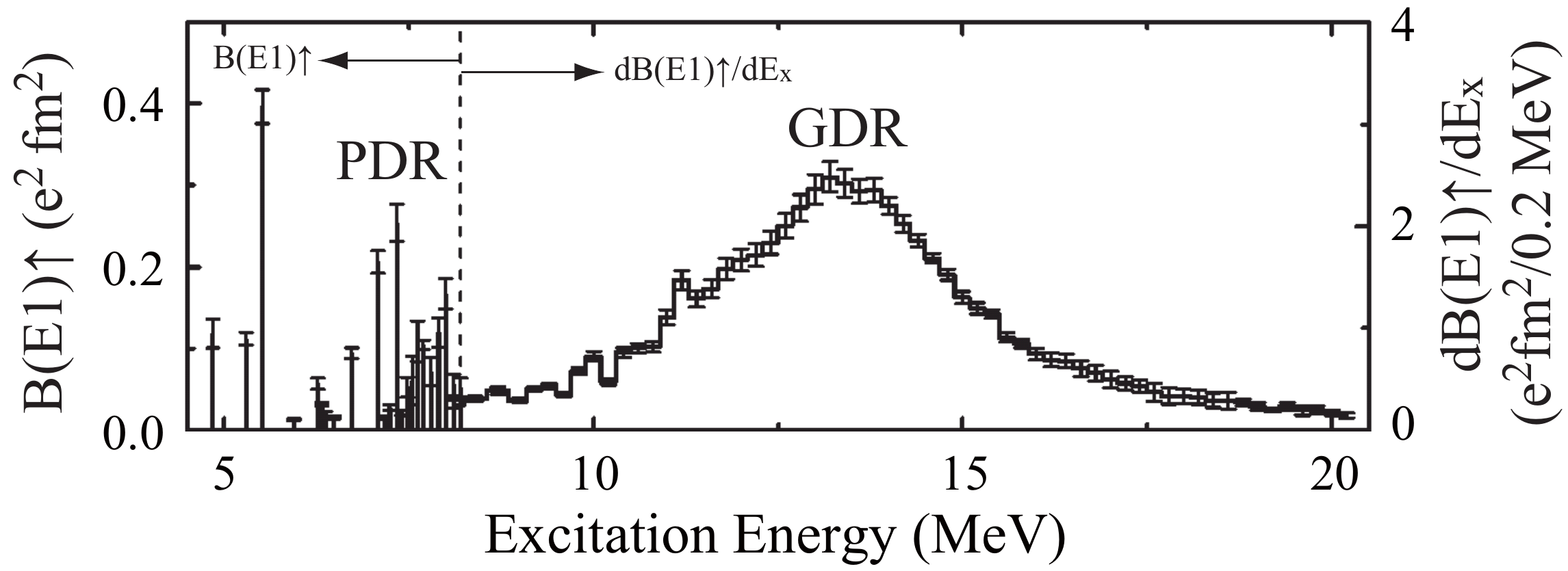}
\caption{\label{fig:BE1}
Total $B(E1)$ strength distribution of $^{208}{\rm Pb}$ deduced from the present work.
The bump centered at $\sim$13 MeV corresponds to the giant dipole resonance, and the strength concentration at around 7-9 MeV to the pygmy dipole resonance.
}
\end{center}
\end{figure}

The overall $B(E1)$ distribution determined by the $(p,p')$ measurement is shown in Fig.~\ref{fig:BE1}.
The bump centered at $\sim$13 MeV corresponds to the GDR and the strength concentration around $7-9$ MeV corresponds to the pygmy dipole resonance (PDR).
A complete $B(E1)$ strength distribution of $^{208}{\rm Pb}$ has been determined from 5 to 20 MeV which fully covers the PDR and GDR regions, as well as region just above  neutron separation energy, where all previous experiments had limited sensitivity.

\section{Dipole polarizability and neutron skin thickness in $^{208}$Pb}
\label{sec:results}

The obtained $B(E1)$ strength distribution is integrated with the aid of Eq.~(\ref{eq:dp}) 
and yields a dipole polarizability $\alpha_D=18.9\pm$1.3 fm$^3$ up to 20 MeV.
By taking the average over the independent data sets and by including 
the gamma absorption data above 20 MeV~\cite{Schelhaas1988}, 
the dipole polarizability of $^{208}{\rm Pb}$ up to 130 MeV is determined as $\alpha_D =20.1\pm0.6$ fm$^3$.

Reinhard and Nazarewicz have shown \cite{Reinhard2010} that in mean-field models, 
for variations of the interaction within reasonable limits there is a strict correlation 
between the predictions for the dipole polarizability and neutron skin thickness of $^{208}{\rm Pb}$.
With the particular interaction of Ref.~\cite{Reinhard2010} one can derive from the above experimental 
result for $\alpha_D$  a value $\Delta r_{np}=0.156^{+0.025}_{-0.021}$ fm \cite{Tamii2011}.
%
\begin{figure}[t]
\begin{center}
\includegraphics[width=20pc]{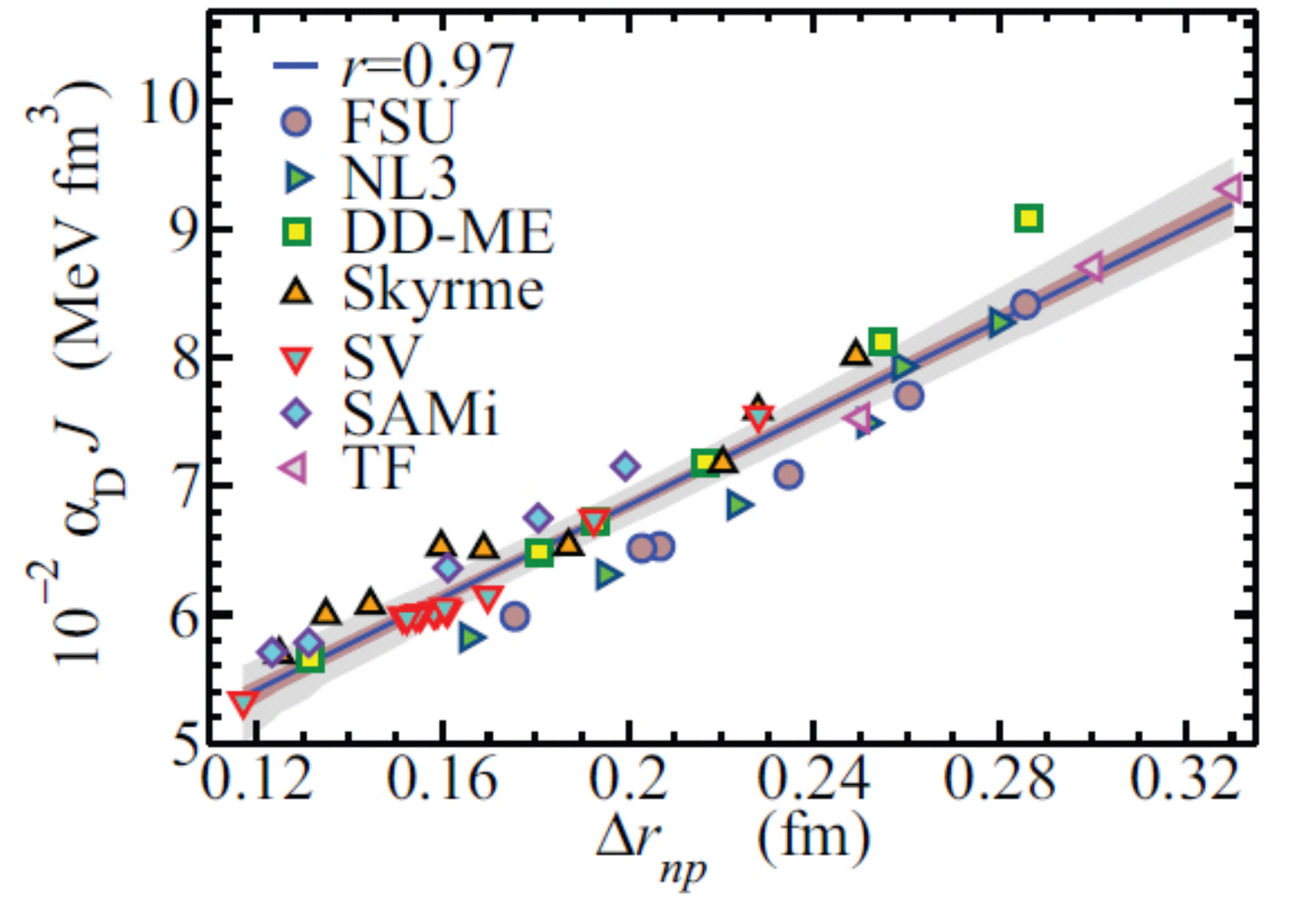}
\caption{\label{fig:NST-DPJ}
Correlation between the dipole polarizability $\alpha_D$ times $J$ and the neutron skin thickness $\Delta r_{np}$ 
in $^{208}$Pb for various mean-field models. 
Figure reprinted with permission from X.~Roca-Maza {\it et al.}~\cite{Roca-Maza2013}.
Copyright (2013) by the American Physical Society.
}
\end{center}
\end{figure}
Piekarewicz {\it et al.}~\cite{Piekarewicz2012} have studied this correlation for a variety 
of non-relativistic and relativistic density functionals (EDFs).
However, absolute values, while indicating an overall approximately linear correlation, exhibit more 
scattering upon a detailed look.
If averaged over the set of theoretical results falling into the experimental uncertainty of $\alpha_D$, 
one obtains a value for the neutron skin thickness in $^{208}$Pb of $0.168 \pm 0.022$ fm~\cite{Piekarewicz2012}.
Roca-Maza {\it et al.}~\cite{Roca-Maza2013} have further investigated the correlation.
They have found that $\alpha_D J$, instead of $\alpha_D$, has much stronger correlation with $\Delta r_{np}$
(see Fig.~\ref{fig:NST-DPJ}) and the correlation is naturally explained by the macroscopic droplet model.
The value of the neutron skin thickness of $^{208}{\rm Pb}$ as a function of $J$ is, by using the experimental
value of $\alpha_D$
\begin{eqnarray}
\Delta r_{np}&=&-0.157\pm(0.002)_{\rm theor}\nonumber\\
&&+[1.04\pm(0.03)_{\rm expt}\pm(0.04)_{\rm theor}]\times10^{-2} J,
\end{eqnarray}
where $\Delta r_{np}$ is expressed in fm and $J$ in MeV.
The ``expt'' uncertainty refers to the propagation of the experimental uncertainty of $\alpha_D$, whereas
the ``theor'' uncertainties are associated with the confidence bands from the theoretical linear fit.
Adopting $J=[31\pm(2)_{\rm est}]$ MeV as a realistic range of values for the symmetry energy~\cite{Tsang2012,LattimerLim2013},
they extracted the constraint on the neutron skin thickness of $^{208}{\rm Pb}$ as~\cite{Roca-Maza2013}
\begin{equation}
\Delta r_{np} = 0.165\pm(0.009)_{\rm expt}\pm(0.013)_{\rm theor}\pm(0.021)_{\rm est}\,{\rm fm},
\end{equation}
where ``est'' uncertainty is associated with the estimates on $J$.

\begin{figure}[h]
\begin{center}
\includegraphics[width=20pc]{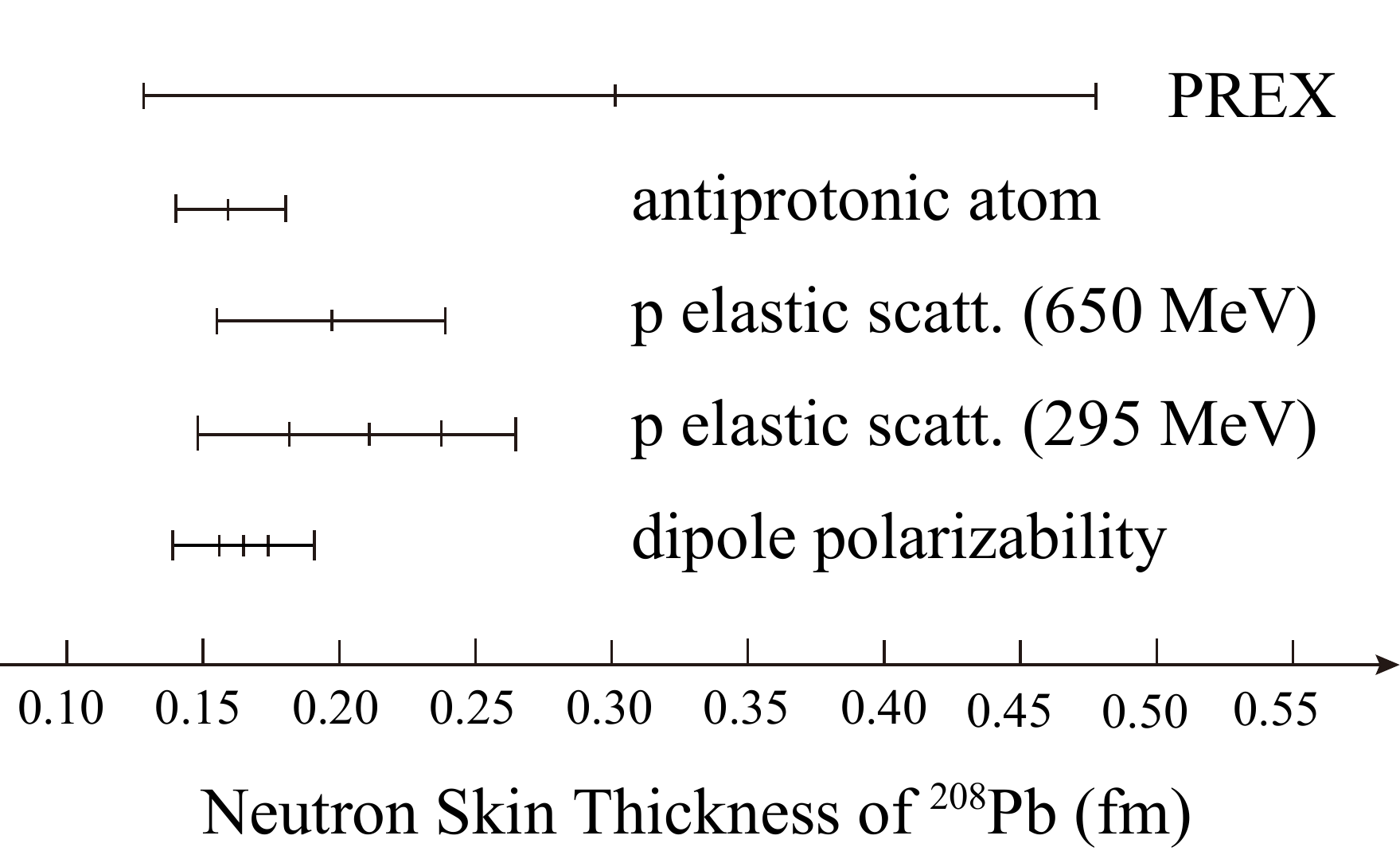}
\caption{\label{fig:NST}
The neutron skin thickness of $^{208}{\rm Pb}$ measured by parity-violating
electron scattering (PREX)~\cite{PREX,PREXResult},
analysis of the X-ray detection data of anti-protonic atoms~\cite{Klos2007},
proton elastic scattering at 650 MeV~\cite{elastic650} and 295 MeV~\cite{Zenihiro2010},
and the dipole polarizability (this work)~\cite{Tamii2011,Roca-Maza2013}.
}
\end{center}
\end{figure}

This result is compared with other experimental work in Fig.~\ref{fig:NST}.
The inner error bar of the dipole polarizability analysis refers to the experimental uncertainty (0.009 fm)
and the outer error bar to the quadratic sum of all the uncertainties (0.026 fm).
As pointed out above, the statistics of the  PREX experiment are presently insufficient to 
provide relevant boundaries on the neutron skin thickness  \cite{PREXResult}.
Antiprotonic atom annihilation and shifts of X rays have been used to derive the neutron density of $^{208}$Pb, and combined with the charge density distribution the neutron skin thickness \cite{Trzcinska2001,Klos2007}. 
Another method is based on extracting the matter radius from elastic proton scattering.
Two results obtained at 650 MeV~\cite{elastic650} and 295 MeV~\cite{Zenihiro2010} are currently available.
All results agree within error bars and limit the range of the neutron skin thickness to $\sim 0.15 - 0.20$ fm.
Recent calculations of neutron matter and neutron star properties in the framework of chiral effective field 
theory suggest $\Delta r_{np}= 0.17 \pm 0.03$ fm \cite{Hebeler2010}. 
The predictions are sensitive to three-nucleon forces, which may be further constrained by the present results.

\section{Constraints on the symmetry energy}
\label{sec:symmetry}

Mean-field models predict an approximately linear relation between the neutron skin thickness of $^{208}$Pb 
and the slope parameter $L$ of the symmetry energy~\cite{Roca-Maza2011,Brown2000}. 
Thus one can also expect a linear relation between $\alpha_D J$ and $L$.
Actually Roca-Maza {\it et al.}~\cite{Roca-Maza2013} have shown a strong correlation 
between $\alpha_D J$ and $L$.~\cite{Roca-Maza2013}
They have extracted the relation by using the dipole polarizability data as
\begin{equation}
L=-146\pm(1)_{\rm theor}+[6.11\pm(0.18)_{\rm expt}\pm(0.26)_{\rm theor}] J,
\end{equation}
and by adopting $J=[31\pm(2)_{\rm est}]$ MeV
\begin{equation}
L=43\pm(6)_{\rm expt}\pm(8)_{\rm theor}\pm(12)_{\rm est}\,{\rm MeV}.
\end{equation}


Since constraints on the symmetry energy parameters $J$ and $L$ are the main concern in this paper,
we have extracted a constraint band in the $J$-$L$ plane from the dipole polarizability data
without making assumption on the $J$ value. We have taken the quadratic sum of the 99.9\% theoretical 
confidence level band from the work by Roca-Maza {\it et al.}~\cite{Roca-Maza2013} and the one-sigma 
experimental uncertainty of $\alpha_D$~\cite{Tamii2011}.
Within the applicable range of $24<J<42$ MeV, the band is placed in between the two curves of
$L=-93.5+3.22J+0.0461J^2$ and $L=-182.4+7.85J-0.0266J^2$.
If the two-sigma experimental uncertainty of $\alpha_D$ is used, the constraint band is
in between 
$L=-111.0+4.43J+0.0307J^2$ and $L=-172.0+7.12J-0.0189J^2$.

\begin{figure}[h]
\begin{center}
\includegraphics[width=20pc]{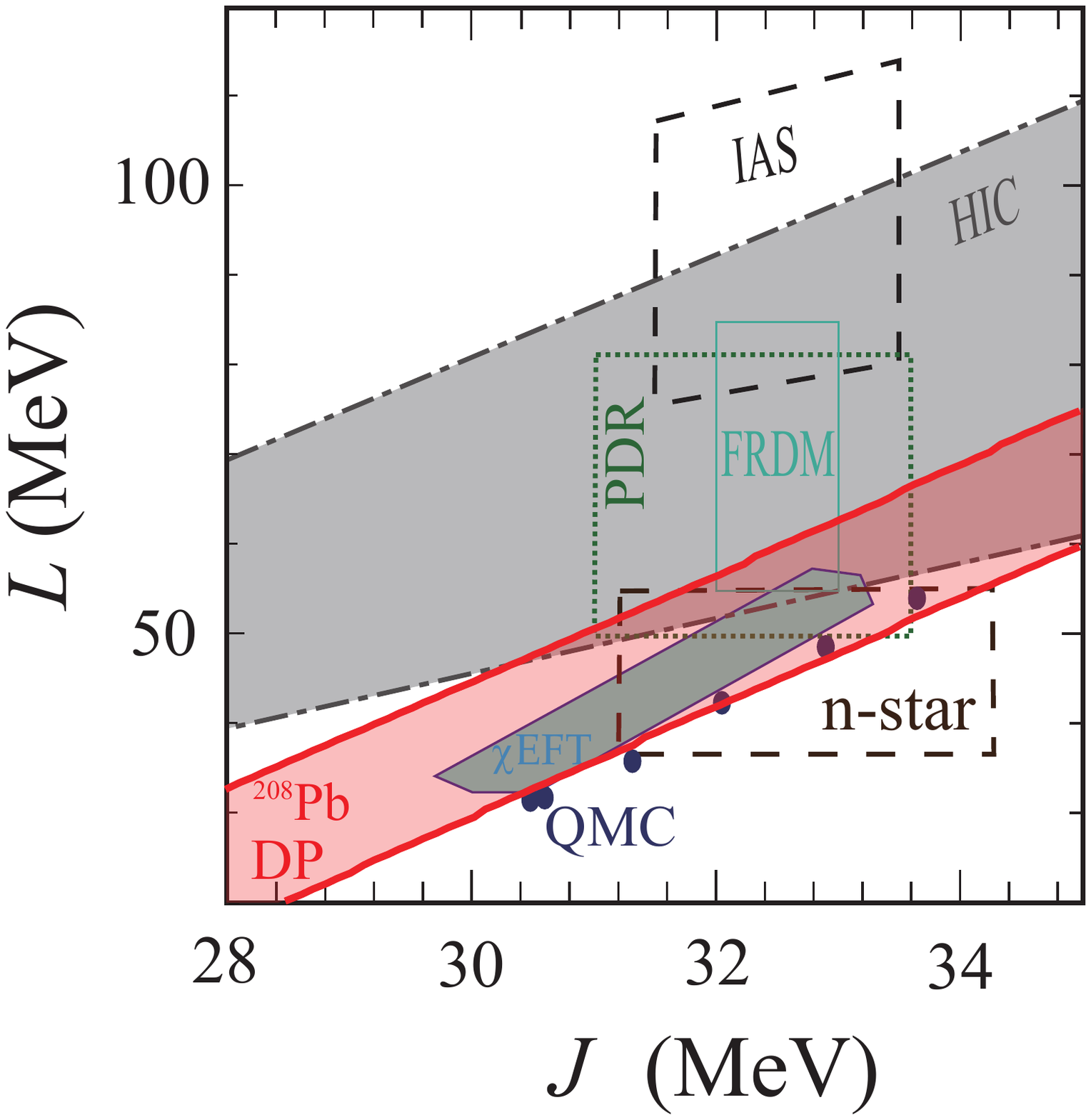}
\caption{\label{fig:J-L}
Constraints on the symmetry energy parameters $J$ and $L$ from various methods.
Besides the present work based on the $^{208}$Pb dipole polarizability (DP), the results in this figure are taken from Ref.~\cite{Tsang2012} except the ones labeled ($\chi$EFT) from Ref.~\cite{Hebeler2010} and QMC (solid circles) from Ref.~\cite{Gandolfi2013}.
}
\end{center}
\end{figure}

The result is plotted in Fig.~\ref{fig:J-L} for the one-sigma uncertainty 
in comparison with various constraints from other works.
These include heavy ion collisions (HIC), pygmy dipole resonance (PDR), isobaric analog states (IAS), 
nuclear mass formula with finite range droplet model (FRDM), analysis of neutron star observation data (n-star).
For a thorough discussion of these methods and the corresponding references see Ref.~\cite{Tsang2012}.
Additionally, constraints from a chiral effective field theory ($\chi$EFT) calculation including 3$N$ 
forces~\cite{Hebeler2010}
and a quantum Monte-Carlo (QMC) calculation~\cite{Gandolfi2013} are shown.
One finds a point ($L \approx 50$ MeV, $J \approx 32$ MeV) essentially consistent with all approaches 
except for the study of isobaric analogue states which requires much higher values of $L$.
Indeed, these common values are at the center of the allowed $L,J$ landscape 
in the first complete N$^3$LO calculation of neutron matter within $\chi$EFT \cite{Tews2013}.

We note that in the paper of Lattimer~\cite{Lattimer2013} an anticorrelation 
between the $L$ and $J$ is deduced from $\alpha_D$ of $^{208}$Pb based 
on a Skyrme-Hartree-Fock approach~\cite{Chen2010} in contradiction to the present result.
The anticorrelation originates from a two step evaluation of the constraint in the $J$-$L$ plane
from $\alpha_D$ by way of $\Delta r_{np}$, whereas the present result is more direct.

\section{Summary and outlook}
\label{sec:summary}

The electric dipole response of ${}^{208}{\rm Pb}$ has been determined up to 20 MeV by proton inelastic scattering measurement at extreme forward angles.
Multipole components of the cross sections have been decomposed by two independent methods: polarization transfer analysis and angular distribution analysis.
The dipole polarizability of $^{208}{\rm Pb}$ up to 130 MeV was determined as $\alpha_D=20.1\pm0.6$ fm$^3$ 
by combining the present and other available data.
This precise result allows to constrain the neutron skin thickness of $^{208}{\rm Pb}$ 
to $\Delta r_{np}=0.165\pm(0.009)_{\rm expt}\pm(0.013)_{\rm theor}\pm(0.021)_{\rm est}$ fm.
A constraint band has been extracted for the symmetry energy and its slope parameter 
at the saturation density utilizing the correlations between $\alpha_D J$ and $L$ obtained by 
a study of mean-field models in EDF approach.
The accuracy already sets important constraints for model parameters of EDFs 
and predictions of neutron star properties, supernova explosion dynamics, 
and many other interesting astrophysical phenomena.

In order to further constrain the symmetry energy parameters -- or more precisely EDF approaches, from which they can be derived --,  dipole polarizability data for other nuclei would be important. 
A particularly interesting candidate is $^{48}{\rm Ca}$, for which theoretical predictions (especially with Skyrme-type forces) of the neutron skin thickness are more scattered and relatively uncorrelated with that of $^{208}{\rm Pb}$ \cite{Piekarewicz2012}.
Proton inelastic scattering data on $^{48}{\rm Ca}$ are already measured.
Analysis and extraction of the dipole polarizability for $^{48}{\rm Ca}$ is in progress as well as for $^{90}{\rm Zr}$~\cite{Iwamoto2012}, $^{96}{\rm Mo}$, $^{120}{\rm Sn}$, $^{144,154}{\rm Sm}$.

Beyond the extraction of the dipole polarizability, we note that the experimental data provide information on the PDR strength distribution~\cite{Poltoratska2012,Iwamoto2012}, the spin-M1 strength distribution~\cite{Heyde2010,Birkhan2013}, fine structure of the GDR and its interpretation in terms of characteristic scales \cite{Shevchenko2004}, level density of the $E1$ strength using a fluctuation analysis~\cite{Kalmykov2006,Kalmykov2007}, and the gamma strength function.

\section*{Acknowledgments}
\label{sec:acknowledgments}

This work was performed by the RCNP-E282 collaboration and we thank all coauthors of Refs.~\cite{Tamii2011,Poltoratska2012} for their contributions.
We are indebted to the RCNP cyclotron accelerator staff and operators for providing an excellent beam.
Discussions with W.~Nazarewicz, J.~Piekarewicz, P.-G.~Reinhard, A.~Schwenk, and X.~Roca-Maza are 
gratefully acknowledged.
This work was supported by JSPS (Grant Nos. 07454051 and 14740154) and 
DFG (Contracts SFB 634 and NE 679/3-1).


\begin{thebibliography}{}
\bibitem{Tsang2012}
\label{ref:Tsang2012}
M.B.~Tsang {\it et al.}, {\it Phys. Rev. C} {\bf 86}, 015803 (2012).
\bibitem{Horowitz2001}
C.J.~Horowitz, J. Piekarewicz, {\it Phys. Rev. Lett.} {\bf 86}, 5647 (2001).
\bibitem{Lattimer2007}
J.M.~Lattimer and M. Prakash, {\it Phys. Rep.} {\bf 442}, 109 (2997).
\bibitem{Roca-Maza2011}
X.~Roca-Maza, M. Centelles, X. Vi\~{n}as, M. Warda, {\it Phys. Rev. Lett.} {\bf 106}, 252501 (2011).
\bibitem{Piekarewicz2012}
J.~Piekarewicz {\it et al.}, {\it Phys. Rev. C} {\bf 85}, 041302(R) (2012).
\bibitem{PREX}
S. Abrahamyan {\it et al.}, {\it Phys. Rev. Lett.} {\bf 108}, 112502 (2012).
\bibitem{PREXResult}
C.J.~Horowitz {\it et al.}, {\it Phys. Rev. C} {\bf 85}, 032501(R) (2012).
\bibitem{Reinhard2010}
P.-G. Reinhard, W. Nazarewicz, {\it Phys. Rev. C} {\bf 81}, 051303(R) (2010).
\bibitem{Tamii2009}
A. Tamii {\it et al.}, {\it Nucl. Instrum. Meth. in Phys. Res. Sect. A} {\bf 605}, 326 (2009).
\bibitem{Tamii2011}
A. Tamii {\it et al.}, {\it Phys. Rev. Lett.} {\bf 107}, 062502 (2011).
\bibitem{Poltoratska2012}
I. Poltoratska, {\it et al.}, {\it Phys. Rev. C} {\bf 85}, 041304(R) (2012).
\bibitem{GR} 
M. Fujiwara {\it et al.}, {\it Nucl. Instrum. Meth. in Phys. Res. Sect. A} {\bf 422}, 484 (1999).
\bibitem{Ryezayeva2002}
N. Ryezayeva {\it et al.}, {\it Phys. Rev. Lett.} {\bf 89}, 272502 (2002).
\bibitem{Enders2003}
J.~Enders {\it et al.}, {\it Nucl. Phys. A} {\bf 724}, 243 (2003);
\bibitem{Shizuma2008}
T. Shizuma {\it et al.}, {\it Phys. Rev. C} {\bf 78}, 061303(R) (2008). 
\bibitem{Schwengner2010}
R.~Schwengner {\it et al.}, {\it Phys. Rev. C.} {\bf 81}, 054315 (2010).
\bibitem{Ohlsen1972}
G.G.~Ohlsen, {\it Rep. Prog. Phys.} {\bf 35}, 717 (1972).
\bibitem{Suzuki2000}
T. Suzuki, {\it Prog. Theor. Phys.} {\bf 103}, 859 (2000).
\bibitem{Laszewski1998}
R.M. Laszewski, R. Alarcon, D.S. Dale, S.D. Hoblit, {\it Phys. Rev. Lett.} {\bf 61}, 1710(1988).
\bibitem{DWBA07}
J. Raynal, computer code {\it DWBA07}, NEA data bank NEA-1209.
\bibitem{Franey1985} 
M.A. Franey and W.G. Love, {\it Phys. Rev. C} {\bf 31}, 488 (1985).
\bibitem{Veyssiere1970}
A. Veyssiere {\it et al.}, {\it Nucl. Phys. A} {\bf 159}, 561 (1970).
\bibitem{Schelhaas1988}
K.P.~Schelhaas {\it et al.}, {\it Nucl. Phys. A} {\bf 489}, 189 (1988).\bibitem{Trzcinska2001}
A.~Trczi\'nska {\it et al.}, {\it Phys. Rev. Lett.} {\bf 87},  082501 (2001).
\bibitem{Roca-Maza2013}
X.~Roca-Maza, M. Brenna, G. Col\`{o} {\it et al.}, {\it Phys. Rev. C.} {\bf 88}, 024316 (2013).
\bibitem{LattimerLim2013}
J.M.~Lattimer and Y. Lim {\it Astrophys. J.} {\bf 771}, 51 (2013).
\bibitem{Klos2007}
B.~Klos {\it et al.}, {\it Phys. Rev. C} {\bf 76}, 014311 (2007).
\bibitem{elastic650}
V.E.~Starodubsky and N.M.~Hintz, {\it Phys. Rev. C} {\bf 49}, 2118 (1994).
\bibitem{Zenihiro2010}
J.~Zenihiro {\it et al.}, {\it Phys. Rev. C} {\bf 82}, 044611 (2010).
\bibitem{Hebeler2010}
K. Hebeler, J.M. Lattimer, C.J. Pethick, A. Schwenk, {\it Phys. Rev. Lett.} {\bf 105}, 161102 (2010).
\bibitem{Gandolfi2013}
S.~Gandolfi, J.~Carlson, S.~Reddy, W. Steiner, and R.B.~Wiringa {\it et al.}, arXiv:1307.5815 [nucl-th] (2013);
S.~Gandolfi, private communication.
\bibitem{Brown2000}
B.A. Brown, {\it Phys. Rev. Lett.} {\bf 85}, 5296 (2000); S. Typel
and B.A. Brown, {\it Phys. Rev. C} {\bf 64}, 027302 (2001).
\bibitem{Tews2013}
I. Tews, T. Kr\"uger, K. Hebeler, and A. Schwenk, {\it Phys. Rev. Lett.} {\bf 110}, 032504 (2013).
\bibitem{Lattimer2013}
J.M. Lattimer, {\it Annu. Rev. Nucl. Part. Sci.} {\bf 62}, 485 (2012).
\bibitem{Chen2010}
Lie-Wen Chen, Che Ming Ko, Bao-An Li, Jun Xu, {\it Phys. Rev. C} {\bf 82}, 024321 (2010).
\bibitem{Roca-MazaPC}
X.~Roca-Maza, private communication.
\bibitem{Iwamoto2012}
C.~Iwamoto {\it et al.}, {\it Phys. Rev. Lett.} {\bf 108}, 262501 (2012).
\bibitem{Heyde2010}
K.~Heyde, P. von Neumann-Cosel, A. Richter, {\it Rev. Mod. Phys.} {\bf 82}, 2365 (2010).
\bibitem{Birkhan2013} 
J. Birkhan {\it et al.}, submitted; arXiv:1308.2817 [nucl-ex].
\bibitem{Shevchenko2004}
A. Shevchenko {\it et al.}, {\it Phys. Rev. Lett.} {\bf 93}, 122501 (2004).
\bibitem{Kalmykov2006}
Y.~Kalmykov {\it et al.}, {\it Phys. Rev. Lett.} {\bf 96}, 012502 (2006).
\bibitem{Kalmykov2007}
Y.~Kalmykov {\it et al.}, {\it Phys. Rev. Lett.} {\bf 99}, 202502 (2007).
\end{thebibliography}
\end{document}